\documentstyle[12pt,epsf]{article}

\begin{document}

\title{ \Large\bf
 Instantons and the Chiral Phase Transition }

\author{Thomas Appelquist\thanks{Electronic address:
 twa@genesis3.physics.yale.edu}\ \ and
 Stephen B. Selipsky\thanks{Electronic address:
 stephen@genesis1.physics.yale.edu}\\ \\ {\it
 Department of Physics, Yale University, New Haven, CT 06520-8120}}

\date{February 23, 1997}

\maketitle

\begin{picture}(0,0)(0,0)
\put(346,226){hep-ph/9702404}
\put(360,210){YCTP-P1-97}
\end{picture}
\vspace{-24pt}

\begin{abstract}
   We examine the role of instantons in the zero-temperature chiral phase
transition in an $SU(N)$ gauge theory.  For a range of $N_f$ (the number
of fermion flavors) depending on $N$, the theory exhibits an infrared fixed
point at coupling $\alpha_*$.  As $N_f$ decreases, $\alpha_*$ increases,
and it eventually exceeds a critical value sufficient to trigger chiral
symmetry breaking.  For the case $N = 2$, we estimate the critical values
of $N_f$ and $\alpha_*$ due to instantons by numerically solving a gap
equation with an instanton-generated kernel.  We find instanton effects
of strength comparable to that of gluon exchange.
\end{abstract}

   Instanton configurations \cite{BPST} of the Yang-Mills potentials
$A^a_\mu(x)$ have been studied extensively for over two decades.
They play a central role in the solution of the QCD $U(1)$ problem
\cite{tHooft}, and a host of other physical consequences have been
examined \cite{instrefs}.  In particular, many authors have studied their
possible role in the dynamical breaking of chiral symmetry in QCD
\cite{CallenDashenGross, CarlitzCreamer, early-inst-chir, shuryak, diakonov}.
All these studies face a difficulty:  their effects are dominated by large
instantons, on the order of the inverse confinement scale of the theory,
where the interactions become strong and instantons overlap.
Reliable quantitative estimates are therefore difficult.

   A recent paper \cite{ATW96} suggested that the chiral phase transition
in an $SU(N)$ theory at zero temperature, as a function of the number of
fermions $N_f$, could be analyzed without the complications of confinement.
For a certain range of $N_f$, the two-loop $\beta$ function has an infrared
stable fixed point, with the fixed point coupling $\alpha_*$ increasing
as $N_f$ decreases.  The transition was argued to set in when $\alpha_*$
exceeded a certain critical value.  That work considered forces arising
solely from gluon exchange.

   Here, we examine the role of instantons in the same theory.
The fixed point allows a more reliable study of instanton effects because
it limits the growth of the effective coupling at the large length scales
which dominate the dynamics, better controlling the integration over the
size of single instantons.  We will begin our presentation with a brief
review of the model; then display a gap equation with a kernel appropriate
to an instanton background; next qualitatively consider the nature of
possible solutions before displaying numerical results;
and finally discuss reliability and draw some physical conclusions.

   We write the Lagrangian for $SU(N)$ gauge theory as
\begin{equation} \label{lagrangian}
 {\cal L}\ =\ \bar{\psi}\Bigl(
 i\!\not\!\partial - g\not\!\!A^a T^a \Bigr)\psi
 - {1 \over 4} F^a_{\mu\nu} F^{a\mu\nu}
 + \Bigl({\rm gauge\ fixing\ terms}\Bigr)\ ,
\end{equation}
where $g$ is the gauge coupling, $\psi$ describes $N_f$ flavors
of Dirac fermions in the $SU(N)$ fundamental representation,
and the adjoint index $a$ ranges over $(1,\cdots , N^2 - 1)$.
We assume a vanishing $\theta$ parameter multiplying the anomaly term
$F \widetilde F$.

   After renormalization, the coupling $\alpha(\mu) \equiv g^2(\mu) / 4\pi$
runs with energy scale, obeying a renormalization group equation
\begin{equation} \label{betafunction}
 \mu {\partial \over \partial\mu} \alpha(\mu)\ \equiv\ \beta(\alpha)\
 =\ - b\alpha^2(\mu) - c\alpha^3(\mu) - \dots\ .
\end{equation}
The first two coefficients are renormalization scheme independent:
\begin{equation} \label{betab}
 b\ =\ {1 \over 6\pi} \left( 11 N - 2 N_f \right)
\end{equation}
\begin{equation} \label{betac}
 c\ =\ {1 \over 24 \pi^2}
 \left(34 N^2 - 10 N N_f - 3{{N^2 - 1}\over N} N_f \right)\ .
\end{equation}
Asymptotic freedom requires $b > 0$ (or $N_f < 11 N/2$).  If in addition
$c < 0$, there is an infrared stable, non-trivial fixed point
\cite{IRfixpt}, located (to two-loop accuracy) at coupling
\begin{equation} \label{alphastar}
 \alpha_*\ =\ -\, {b\over c}\ .
\end{equation}
Assuming that at some reference scale $\Lambda$, the running coupling has
a value $\alpha(\Lambda)$ between 0 and $\alpha_*$, we then have
$\alpha(\mu \ll \Lambda) \rightarrow \alpha_*$ and
$\alpha(\mu \gg \Lambda) \rightarrow 0$.  We can write the two-loop
solution to the renormalization group Eq.~(\ref{betafunction}) in the form
\begin{equation} \label{run2loop}
 \Lambda /\mu\ =\
 \Bigl|{\alpha_* / \alpha(\mu)} - 1\Bigr|^{-1/(b\alpha_*)}\,
 \exp\Bigl[{-1/b\alpha(\mu)}\Bigr]\ ,
\end{equation}
giving $\alpha(\Lambda) = 0.7822\, \alpha_*$.
Note that $\alpha(\Lambda)$ will vary when we change $N_f/N$.

   The reliability of this result depends on whether higher order terms in
the $\beta$ function can be ignored.  That is guaranteed for sufficiently
small $\alpha_*$; more generally, the higher order terms are
renormalization scheme dependent and so can be made arbitrarily small.
For such special renormalization schemes, or for larger $\alpha_*$,
there could of course be important higher order corrections to other
quantities of physical interest \cite{ATW96}.
Nevertheless, the higher-loop $\beta$ function \cite{threeloop} continues
to display an infrared fixed point, with scheme-dependent accuracy
discussed below.

   In Ref.~\cite{ATW96}, it was argued that gluon exchange triggers
dynamical chiral symmetry breaking when the fixed-point coupling
$\alpha_*$ exceeds a critical value
\begin{equation} \label{alphac-gluon}
 \alpha_c\ =\ {\pi\over 3 C_2(R)}\ =\ {2\pi N \over 3(N^2 - 1)}\ .
\end{equation}
This happens when $N_f$ drops below the critical value for gluon exchange,
\begin{equation} \label{Nfc-gluon}
 N_f^{cG}\ =\ N \left(
 {100 N^2 - 66 \over 25 N^2 - 15} \right)\
 \approx\ 4 N (1 - 0.06 N^{-2} - \dots)\ . % - 0.1 (0.6 N^{-2})^n
\end{equation}
The leading approximation for large $N$ is quite good even for $N = 2$.

   An estimate of the higher order corrections to the gap equation driving
the breaking provides some evidence \cite{ALM} for the reliability of these
results.  For $\alpha \leq \alpha_c$, the next order corrections were found
to be relatively small (less than 20\%) compared to the ladder approximation.
As discussed above, there could also be important corrections to the $\beta$
function when $\alpha$ is this large.  Some evidence that this is the case
in the MS scheme adopted here can be provided by determining $\alpha_*$
using the three-loop MS $\beta$ function \cite{threeloop} and setting
this value equal to $\alpha_c$.
For $N = 2$, this leads to a new value of $N_f^{cG}/N \approx 3.175$,
again shifted less than 20\% from the two-loop value 3.929.
The shift decreases to about 15\% as $N \rightarrow \infty$.

   For $N_f < N_f^{cG}$ the gap equation gives a dynamical mass $\Sigma(p)$,
where $p$ is the magnitude of Euclidean momentum.  It has some value
$\Sigma(0)$ at $p = 0$ and then falls monotonically with increasing $p$.
The scale $\Sigma(0)$ vanishes continuously in a characteristic exponential
fashion \cite{ATW96} as $N_f \rightarrow N_f^{cG}$ from below (equivalent
to $\alpha_* \rightarrow \alpha_c$ from above), at fixed $\Lambda$.
Since the fermions decouple for $p \ll \Sigma(p)$, the infrared fixed point
is only an approximate feature of the theory, useful at momentum scales
above the decoupling scale.  Fortunately, the critical behavior of the
theory near the transition is determined mainly by the momentum range
$\Sigma (p) \ll p \ll \Lambda$ in the gap equation integration, where the
fixed point is a good approximation.  Further discussion of the gluon-induced
critical behavior may be found in Refs.~\cite{ATW96, Chivukula, MirYam}.

   Turning now to our study of the role of instantons in the chiral phase
transition, we will derive a corresponding critical coupling or critical
number of flavors $N_f^{cI}$, arising purely from instanton effects.
Comparison with the values from purely gluon exchange will then indicate
the relative importance of the two effects in the phase transition dynamics.

   A nonzero dynamical mass $\Sigma(p)$ in the quark propagator
$i/[A(p)\gamma_\mu p^\mu - \Sigma(p)]$ signals chiral symmetry breaking.
To determine this two-point function we adopt a formalism \cite{CarlitzCreamer}
which self-consistently sums the effects on a fermion propagating through a
dilute gas of noninteracting instantons, giving a gap equation whose kernel
is directly related to the single-instanton amplitude.
For general $N$, the gap equation takes the form
\begin{equation} \label{gapinstanton}
 \Sigma(p)\ =\ \int_0^\infty{d\rho\over\rho^2}\ \Gamma[\alpha(\rho)]\,
 { D[\rho m(\rho)]^{N_f} \over \rho m(\rho) }\, f^2(p\rho /2)
\end{equation}
with
\begin{equation}\label{Gamma}
 \Gamma[\alpha(\rho)]\ =\
 { 4 e^{5/6} e^{C N - B N_f} \over (N - 1)!\, (N - 2)! }
 \left[{2\pi\over\alpha(\rho)} \right]^{2N} e^{-2\pi / \alpha(\rho)}.
\end{equation}
The numerical factors and $\alpha$ dependence in $\Gamma[\alpha(\rho)]$
arise from the amplitude for an instanton of size $\rho$,
integrated over the other collective coordinates \cite{tHooft, Bernard}.
Gauge field and quark fluctuations around the instanton background
contribute quantum corrections which include logarithms that renormalize
the bare coupling to its value at length scale $\rho$.
Depending on the order of the calculation, we end up with the one-
or two-loop renormalization group solution for $\alpha(\rho)$, with
leftover non-logarithmic terms going into the numerical prefactor.

   If the two-loop $\alpha(\rho)$ is to be used, to make use of the
infrared fixed point, then the constants $B$ and $C$ should be computed
to the same order.  They have so far been computed only through one loop
\cite{tHooft, Bernard}, where they are $B = 0.3595$ and $C = 2.0706$
in the MS scheme.\footnote{
 In the MS-bar scheme these become $B = -0.2917$ and $C = -1.5114$.}
The higher-order fluctuations that generate two-loop running in
$\exp{(-2\pi/\alpha)}$ and one-loop running \cite{ABC}
in $\left( 2\pi /\alpha \right)^{2N}$
also contribute corrections of ${\cal O}(\alpha)$ to $B$ and $C$.
We expect that a full two-loop calculation will be scheme independent.
In the absence of such higher order computations,
we will simply take $\alpha(\rho)$ everywhere in the gap equation to be
governed by the two-loop $\beta$ function and infrared fixed point.
For our numerical study, we will use the one-loop, MS values of $B$ and $C$,
assuming that the higher order corrections will lead only to $O(1)$ changes,
in particular keeping $e^B \sim {\cal O}(1)$.

   The function $D[\rho m(\rho)]$ contains the mass-dependent factors from
the fermionic quantum fluctuations (left over after the factors containing
regulated divergences are absorbed into $\Gamma[\alpha(\rho)]$).
Following Ref.~\cite{CarlitzCreamer} we evaluate the argument of
$D[\rho m(\rho)]$ using the function $m(\rho)$ derived from $\Sigma(p)$
weighted by the fermion zero mode wavefunction,
\begin{equation} \label{mtransform}
 m(\rho)\ =\ \langle\psi_0 |\Sigma |\psi_0\rangle \ =\
 \int_0^\infty{ dx\ x\, f^2(x/2)\, \Sigma(p = x/\rho) }\ .
\end{equation}
Here, $f(x)$ is a combination of modified Bessel functions \cite{abramowitz}
arising from the Fourier transform of $\psi_0(x)$,
\begin{equation} \label{fbessel}
 f(x)\ \equiv\ - 2I_1(x)K_1(x) - 2x\Bigl(I_1(x)K_0(x) - I_0(x)K_1(x)\Bigr)\ ,
\end{equation}
normalized to $f(0) = 1$.
Its asymptotic behavior is $f(x\gg 1) \sim {3\over 4}x^{-3}$.

   For arguments $\rho m(\rho) \ll 1$, the fermion fluctuation factor
$D[\rho m(\rho)]$ has the expansion $D(x) = x + {\cal O}(x^3\ln{x})$.
For large arguments, the fermions decouple and $D[\rho m(\rho)]$
exponentially approaches unity \cite{AndreiGross}, which we must multiply
by $e^B \sim {\cal O}(1)$ to account for decoupling the fermion factors in
$\Gamma(\alpha)$.  Lacking a full calculation of the instanton determinant
for massive fermions, we adopt a simple form that interpolates between
these two limits:
\begin{equation} \label{Dmatch}
 D(x)\ \equiv\
 \cases{ x & for $x < e^B$ \cr
         1\cdot e^B & otherwise\ . }
\end{equation}

   Spontaneous chiral symmetry breaking corresponds to the existence of
nonvanishing, energetically preferred solutions to the gap equation
(\ref{gapinstanton}).  We will first qualitatively discuss the existence
of these solutions and then summarize the results of a numerical study.
To begin, we note that a nonvanishing solution $\Sigma(p)$ will have
some finite value $\Sigma(0)$ at $p = 0$, and then decrease monotonically.
This behavior, typical for a dynamical mass in a gauge field theory,
follows from the structure of the gap equation (\ref{gapinstanton})
with the factor $f^2 (p\rho /2)$ decreasing monotonically from unity.
The zero mode mass in Eq.~(\ref{mtransform}) is approximately
$m(\rho) \sim \Sigma(1/\rho)$, since $x f^2(x/2)$ is peaked around $x = 1$
and integrates to unity.  Thus, $m(\rho \rightarrow \infty) = \Sigma(0)$,
while $m(\rho)$ falls rapidly as $\rho \rightarrow 0$.  Although the
intrinsic scale $\Lambda$ sets the solution scale $\Sigma(0)$, our
interest here is in exploring a possible second order phase transition
near which the gap equation may dynamically enforce $\Sigma(0) \ll \Lambda$.

   Assuming this to be the case,
the integration over $\rho$ then breaks naturally into three regimes.
The ultraviolet regime, $0 < \rho < 1/\Lambda$, contributes very little
to the integral because asymptotic freedom ensures a strong suppression
of $\Gamma[\alpha(\rho)]$, while $D(\rho m)^{N_f}$ also remains small.
In the intermediate regime, $\Lambda^{-1} < \rho
 {\ \lower-1.2pt\vbox{\hbox{\rlap{$<$}\lower5pt\vbox{\hbox{$\sim$}}}}\ }
\Sigma(0)^{-1}$, $\alpha(\rho)$ ranges only from roughly $0.78 \alpha_*$
to $\alpha_*$, and the infrared fixed point dominates the behavior.
The upper end of this intermediate regime, $\rho \approx \Sigma(0)^{-1}$,
should dominate the integral due to the polynomial increase of
$D[\rho m(\rho)]$ while $\Gamma[\alpha(\rho)]$
simply approaches its fixed point value $\Gamma_*$.
In the third regime, as $\rho \gg \Sigma(0)^{-1}$ the fermions decouple
from the fluctuations, ending the polynomial increase of $D[m\rho]$ and
the fixed point behavior of $\alpha$.
The third regime will affect the critical value of $N_f$ but not the
qualitative behavior near criticality.

   We can see this by scaling to dimensionless variables
$s(y) = \Sigma(0)^{-1} m(y/\Sigma(0))$ with $y = \Sigma(0) \rho$;
note that $0 \le s(y) \le 1$.  The gap equation at $p = 0$ then becomes
\begin{equation} \label{gapqualitative}
 1\ \approx\ 0 \ +\ \Gamma_* \int_{\Sigma(0)/\Lambda}^1
  { {dy\over y^2}\, [y s(y)]^{N_f - 1} }
 \ +\ \int_1^\infty{ \Gamma(y)\, e^{BN_f}\, {dy\over y^3 s(y)} }
\end{equation}
for nonvanishing $\Sigma(0)$.  Thus it is the intermediate regime that
controls the critical behavior of $\Sigma(0)/\Lambda$.  An important
feature of Eq.~(\ref{gapqualitative}) is that since $s(y) \le 1$,
it cannot be satisfied if $\Gamma_* \ll 1$, which occurs for $N_f$ close
enough to $11 N/2$ (that is, small $\alpha_*$).  Only the chirally symmetric
solution $\Sigma(0) = 0$ exists for this range of $N_f$.

   As we decrease $N_f$ to $N_f^{cI}$ (the critical number of flavors for
the instanton kernel), $\Gamma_*$ will eventually reach a critical value
large enough to allow nonzero solutions.  The structure of
Eq.~(\ref{gapqualitative}) indicates that this will correspond to
maximizing the intermediate integral, and therefore to taking its lower
limit to zero.  For $N_f$ slightly less than $N_f^{cI}$,
$\Sigma(0)/\Lambda$ must be small, indicating a continuous phase
transition.  We expect $N_f^{cI}
 {\ \lower-1.2pt\vbox{\hbox{\rlap{$>$}\lower5pt\vbox{\hbox{$\sim$}}}}\ }
3.7 N$, since if a critical value occurs it can only be before $\Gamma_*$
reaches its maximum as a function of $N_f$,
corresponding to $\alpha_* \approx \pi/N$.
We will not here determine analytically the behavior of $\Sigma(0)$
as $N_f \rightarrow N_f^{cI}$.  Instead, having qualitatively seen
that there exists a continuous phase transition at a critical value
$N_f^{cI}$, we now turn to a numerical study of the transition.

   The numerical results reported in this letter are restricted to the
case $N = 2$.  We solve Eq.~(\ref{gapinstanton}) on a one dimensional lattice,
iteratively relaxing the discretized $\Sigma(p)$ from an initial guess to a
self-consistent shape, at each stage numerically integrating to get $m(\rho)$.
We expect, from our qualitative discussion, that the dominant range of
integration will be approximately $\Lambda^{-1} < \rho < \Sigma(0)^{-1}$.
We use the exact solution Eq.~(\ref{run2loop}) to the two-loop
renormalization group equation,
for all $\rho < \Sigma(0)^{-1}$ (more precisely, for $\rho$ below the
value at which $D(\rho m) = e^B$, solved for at each iteration),
using the MS value $B = 0.3595$.
For $\rho$ above this fermion-decoupling value, we match $\alpha(\rho)$
onto the $\beta$ function solution for $N_f = 0$.
After $\alpha$ finally grows too large for perturbative running,
we simply fix it at a constant value $\alpha_{\rm max} = 2\pi/N$;
the far end of the infrared range of integration is safely subdominant,
and we have checked that this approximation is unimportant.

   The result is a shape $\Sigma(p)$ for each value of $N_f$.
To study a possible phase transition we examine the behavior of
$\Sigma(0)$, for fixed $\Lambda$, while varying $N_f$.
The numerical results confirm the qualitative discussion above:  when $N_f$
approaches the critical value $N_f^{cI} \approx 4.77 N \approx 9.54$ from
below, $\Sigma(0)$ vanishes continuously in the manner of a second order
phase transition.  Fig.~1a displays this behavior for $N = 2$;
in contrast to the exponential behavior arising from gluon exchange,
Fig.~1b indicates a power law behavior.

\vskip 1em
\vbox{
\noindent $\log_{10}{[\Sigma(0) / \Lambda]}$
   \hfill $\log_{10}{[\Sigma(0) / \Lambda]}$ \newline
\hbox{ \epsfxsize = 3in \epsfbox{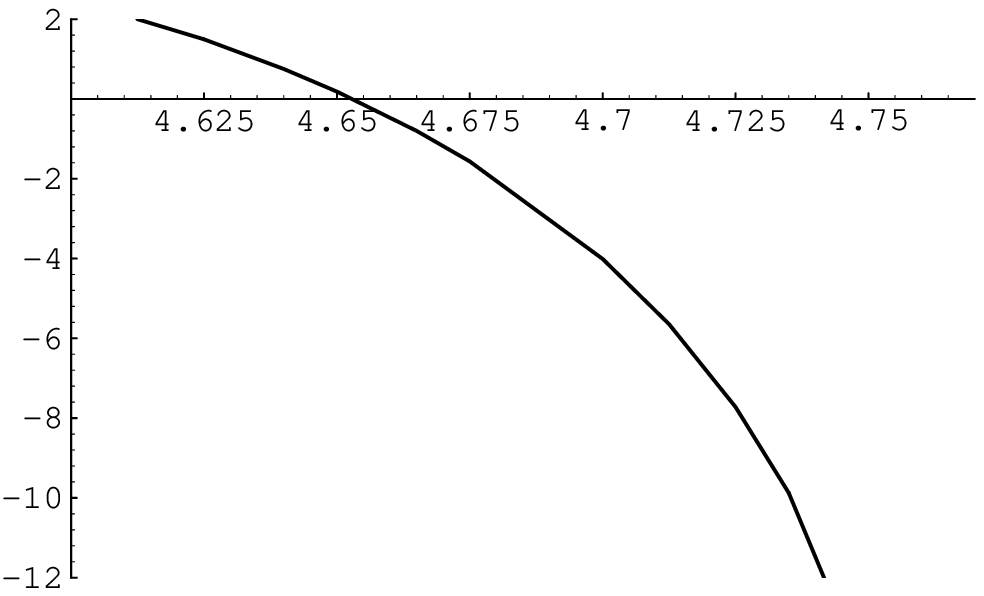}
       \hskip 0.15in %\hfill
       \epsfxsize = 3in \epsfbox{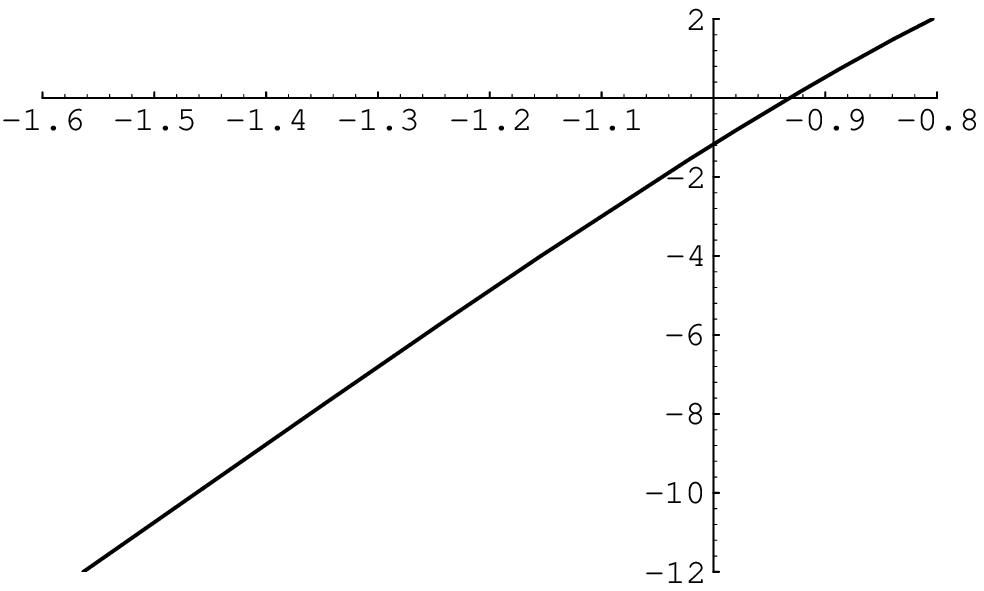}
     }
}\vskip 1em
\vbox{
\noindent{\bf Figure 1}
 (a): $\log_{10}{(\Sigma(0) / \Lambda)}$ against $N_f / N$ for fixed
 $N = 2$.  Extrapolation identifies       \newline \indent\hskip 4.5em
 the critical value $N_f^{cI} = 4.77 N$.  \newline \indent\hskip 2.5em
 (b): Plotting the same data against $\log_{10}{(N_f - N_f^{cI})/N}$
 shows that $\Sigma(0)$                   \newline \indent\hskip 4.5em
 follows a power law in $(N_f - N_f^{cI})$.
}\vskip 1em

   The critical coupling and $N_f^{cI}$ for an instanton kernel are
numerically very similar to the values for the gluon-exchange kernel,
basically independent of whatever ${\cal O}(1)$ values are chosen for
$B$ and $C$.  In any case, a more complete calculation which combined
the two kernels would lead to a somewhat smaller combined critical
coupling (or larger critical $N_f$); the qualitative point is that
the two effects are rather similar in magnitude.

   Finally, we discuss the validity of the dilute gas approximation,
incorporated in the gap equation used here to describe the phase transition.
It allows the amplitude for a fermion propagating in the field of a single
instanton to be summed over multi-instanton configurations, neglecting
instanton interactions.
The validity of this approximation depends on the relative magnitude of the
dominant instanton size and the typical separation distance between instantons
in multi-instanton configurations.  As noted above, the dominant instanton
size is of order $\Sigma(0)^{-1}$, which grows without bound near the
critical point.  But the average instanton separation does just the same,
since it is controlled by essentially the same instanton amplitude integral,
also dominated near $\rho \sim \Sigma(0)^{-1}$.
Crudely estimating the instanton density by
\begin{equation} \label{idensity}
 {\bar n}\ \sim\ \int_0^\infty { {d\rho\over \rho^5}\,
 {\Gamma[\alpha(\rho)] \over 2\pi^2} D[\rho m(\rho)]^{N_f}\ \sim\
 \Sigma(0)^4 \, { e^{BN_f} \Gamma_* \over 2\pi^2 }
 }\ ,
\end{equation}
we ask that there be fewer than one instanton per instanton
four-volume $\Sigma(0)^{-4}$ :
\begin{equation} \label{noverlap}
 {2e^{5/6} \over \pi^2} {e^{CN} \over (N - 1)!\, (N - 2)! }
 \Biggl[{2\pi\over\alpha_*} \Biggr]^{2N} e^{-2\pi / \alpha_*}\ \ll 1 \ .
\end{equation}
For $N = 2$ this requires $\alpha_c$ only slightly smaller than $\pi/N$,
that is $N_f^{cI} > 26 N / (7 - N^{-2}) \approx 3.7 N$.
Our numerical result for $N = 2$ is on the safe side of this limit,
giving some reassurance that instanton overlap does not violate the
dilute gas approximation.  For larger $N$, the non-overlap condition
(\ref{noverlap}) becomes
\begin{equation} \label{largeNoverlap}
 {e^{5/6} \over \pi^3} N^2 \Biggl [ e^{C/2} {2\pi\over N\alpha_*}\,
 e^{1 - \pi / N\alpha_*}\Biggr ]^{2N}\ \ll 1 \ ,
\end{equation}
putting a somewhat stricter bound on $\alpha_{cI}$ and $N_f^{cI}$.

   It is clear from this last expression that at large $N$ the validity
of the dilute gas approximation becomes a delicate matter depending
sensitively on the value of $\alpha_*$, and also on $B$ and $C$.
If $C$ were neglected in Eq.~\ref{largeNoverlap}, the remaining expression
\cite{largeNshuryak} would increase exponentially with $N$,
for $\alpha_*$ in the range required to trigger chiral symmetry breaking.
Since the same factor appears in the instanton amplitude entering the gap
equation, the increase would also affect the dynamics of chiral symmetry
breaking.  That is, the effect of instantons would grow with $N$, at
least up to the point where the dilute gas approximation breaks down.
Whether this actually happens depends on the prefactor term involving
$B$ and $C$.  Until these constants are computed through two loops and
demonstrated to be scheme independent for $\alpha(\rho) = \alpha_*$,
the relevance of instantons to chiral symmetry breaking remains uncertain
in the large $N$ limit.

   In summary, we have studied the role of instantons in the
zero-temperature chiral phase transition in $SU(N)$ gauge theories,
using the number $N_f$ of fermion flavors as the control parameter.
The key feature of these theories is that for a range of $N_f$ including
the critical value for the transition, the two-loop $\beta$ function
exhibits an infrared fixed point.
This allowed us to discuss qualitatively the existence and behavior of
solutions to a gap equation, whose kernel arose from the propagation of
fermions in an instanton background in the dilute gas approximation,
and to present numerical solutions.
We found a critical value of $N_f$, below which chiral symmetry breaking
occurs, which is comparable to that generated by gluon exchange alone.
We conclude that for small $N$, instantons play a role comparable
to that of gluon exchange in the chiral phase transition.

\vskip 3em
%\vfil
   The authors thank Nick Evans, Stephen Hsu and Myckola Schwetz
for useful discussions and comments.
SBS acknowledges the hospitality of the Benasque Center for Physics,
in whose stimulating atmosphere some of this research took place.
This work was supported in part under United States Department of Energy
contract DE-AC02-ERU3075.
%\vfil
\vskip 4em
%\newpage


\begin{thebibliography}{99}

\setlength{\baselineskip}{2ex}
\setlength{\parskip}{0.5ex}

\bibitem{BPST} % Pseudoparticle solutions of the Yang-Mills equations,
A.A. Belavin, A.M. Polyakov, A.S. Shvarts and Yu.S. Tyupkin,
 Phys.\ Lett.\ B59, 85 (1975).

\bibitem{tHooft}
G. 't Hooft, Phys.\ Rev.\ D14, 3432 (1976),
erratum Phys.\ Rev.\ D18, 2199 (1978),
further corrections Phys.\ Rept.\ 142, 357 (1986).

\bibitem{instrefs} % Correlation functions in the QCD vacuum,
E.V. Shuryak, Rev.\ Mod.\ Phys.\ 65, 1 (1993)
 gives a recent review with references to the literature.

\bibitem{CallenDashenGross}
C.G. Callen, R.F. Dashen and D.J. Gross, Phys.\ Rev.\ D17, 2717 (1977).

\bibitem{early-inst-chir}
% Quark mass generation by pseudoparticles,
D.G. Caldi, Phys.\ Rev.\ Lett.\ 39, 121 (1977);
\\
% Bound states from instantons,
R.D. Carlitz, Phys.\ Rev.\ D17, 3225 (1978);
\\
%Propagation functions in pseudoparticle fields, RLO-1388-735, Oct.~1977, 62pp,
L.S. Brown, R.D. Carlitz, D.B. Creamer and Choonkyu Lee,
 Phys.\ Rev.\ D17, 1583 (1978),
% Propagators in pseudoparticle fields, RLO-1388-728, Jun.~1977, 9pp,
 Phys.\ Lett.\ B70, 180 (1977) and Phys.\ Lett.\ B71, 103 (1977);
%(* Published twice in same journal *)
\\
% Instantons and chiral symmetry breaking,
C.E.I. Carneiro and N.A. McDougall, Nucl.\ Phys.\ B245, 293 (1984);
\\
% Instanton Vacuum in Thermal QCD,
V.V. Khoze and A.V. Yung, Z.\ Phys.\ C50, 155 (1991);
% Instanton Vacuum in the Nonzero Temperature QCD, preprint Leningrad-90-1625;
\\
% Asymptotic behavior of quark masses induced by instantons,
C.E.I. Carneiro and J. Frenkel, Phys.\ Rev.\ D31, 949 (1985).

\bibitem{CarlitzCreamer} % Light quarks and instantons,
R.D. Carlitz and D.B. Creamer, Ann.\ Phys.\ 118, 429 (1975).

\bibitem{shuryak} E.V. Shuryak, Novosibirsk preprint INP 87-97 (1987).

\bibitem{diakonov} % Chiral symmetry breaking by instantons,
Dmitri Diakonov, hep-ph/9602375, and references therein.
%Talk given at International School of Physics, 'Enrico Fermi', Course 80:
%Selected Topics in Nonperturbative QCD, Varenna, Italy, 27 June - 7 July 1995.

\bibitem{ATW96}
%The zero temperature chiral phase transition in SU(N) gauge theories,
T. Appelquist, J. Terning and L.C.R. Wijewardhana, hep-ph/9602385,
 Phys.\ Rev.\ Lett.\ 77, 1214 (1996).

\bibitem{IRfixpt}
% Asymptotic behavior of nonabelian gauge theories to two loop order,
W.E. Caswell, Phys.\ Rev.\ Lett.\ 33, 244 (1974);
\\
% Two loop diagrams in Yang-Mills theory,
D.R.T. Jones, Nucl.\ Phys.\ B75, 531 (1974);
\\
D.J. Gross, in {\it Les Houches 1975: Methods in Field Theory,}
 eds. R. Balian and J. Zinn-Justin (North-Holland, 1976, 1981), p.~196;
\\
% On the phase structure of vectorlike gauge theories with massless fermions,
T. Banks and A. Zaks, Nucl.\ Phys.\ B196, 189 (1982).

\bibitem{threeloop} % The four loop beta function in quantum chromodynamics,
T. van Ritbergen, J.A.M. Vermaseren and S.A. Larin, hep-ph/9701390;
\\
% The three loop QCD beta function and anomalous dimensions,
S.A. Larin and J.A.M. Vermaseren, hep-ph/9302208,
 Phys.\ Lett.\ B303, 334 (1993);
\\
% The Gell-Mann-Low function of QCD in the three loop approximation,
O.V. Tarasov, A.A. Vladimirov and A.Yu.\ Zharkov,
 Phys.\ Lett.\ B93, 429 (1980).

\bibitem{ALM} % On the ladder approximation for spontaneous symmetry breaking,
T. Appelquist, K. Lane and U. Mahanta, Phys.\ Rev.\ Lett.\ 61, 1553 (1988).

\bibitem{Chivukula} R.S. Chivukula, hep-ph/9612267.
% A comment on the zero temperature chiral phase transition
%  in SU(N) gauge theories

\bibitem{MirYam} V. Miransky and K. Yamawaki, hep-th/9611142.
% Conformal phase transition in gauge theories

\bibitem{Bernard} C. Bernard, Phys.\ Rev.\ D19, 3013 (1979).
% Gauge zero modes, instanton determinants, and QCD calculations

\bibitem{ABC} % ABC of instantons,
A.I. Vainshtein, V.I. Zakharov, V.A. Novikov and M.A. Shifman,
 Sov.\ Phys.\ Usp.\ 25 (4), 195 (1982).
% Translated from ITEP-2-1981-mc (microfiche), 1981, 84pp.

\bibitem{abramowitz}
M. Abramowitz and I. Stegun, ed.~{\it Handbook of Mathematical Functions}
 (Dover, 1972), eqs.~9.7.1-2, p.~377.

\bibitem{AndreiGross} % The effect of instantons on
% the short distance structure of hadronic currents,
N. Andrei and D.J. Gross, Phys.\ Rev.\ D18, 468 (1978), appendix.

\bibitem{largeNshuryak}
E.V. Shuryak, hep-ph/9503467, Phys.\ Rev.\ D52, 5370 (1995).
% Instanton size distribution: repulsion or the infrared fixed point?

\end{thebibliography}
\end{document}